\documentclass[aps,pra,twocolumn,amsmath,amssymb]{revtex4}
\usepackage{amssymb}
\usepackage{graphicx}
\usepackage{dcolumn}
\usepackage{verbatim}
\usepackage{color,ulem,soul}

\begin{document}

\title{Observation of Three-Mode Parametric Instability}
\author{Xu Chen}
\affiliation{School of Physics, University of Western Australia, WA 6009, Australia}
\author{Chunnong Zhao}
\affiliation{School of Physics, University of Western Australia, WA 6009, Australia}
\author{Li Ju}
\affiliation{School of Physics, University of Western Australia, WA 6009, Australia}
\author{Stefan Danilishin}
\affiliation{School of Physics, University of Western Australia, WA 6009, Australia}
\author{David Blair}
\affiliation{School of Physics, University of Western Australia, WA 6009, Australia}
\author{Haibo Wang}
\affiliation{Department of Physics, Beijing Normal University, Beijing 100875, China}
\author{Sergey P. Vyatchanin}
\affiliation{Faculty of Physics, M.V. Lomonosov Moscow State University, Moscow 119992, Russia}

\date{}

\begin{abstract}

Three-mode parametric interactions can occur in triply-resonant opto-mechanical systems in which two orthogonal optical modes are coupled with an appropriate mechanical mode. Using an optical cavity with a membrane inside, we report the first observation of three-mode parametric instability in a free space Fabry-Perot cavity, a phenomenon predicted to occur in long baseline advanced gravitational wave detectors. We present a large signal model for the phenomenon, which predicts exponential growth of mechanical oscillation followed by saturation. Our experimental results are consistent with this model. Contrary to expectations, parametric instability does not lead to loss of cavity lock, a fact which may make it easier to implement control techniques in Advanced gravitational wave detectors.
\end{abstract}

\maketitle
% \section{Introduction}
In 2001, Braginsky et al. \cite{Braginsky2001, Braginsky2002} predicted three-mode parametric instability (PI) in gravitational waves detectors. The PI would be caused by the coincident match of the mode shapes of mirror mechanical modes and Lageurre-Gauss optical modes. The high mechanical and optical mode density of these systems would ensure that such interactions would occur accidentally. Subsequently detailed modeling \cite{zhao2005} verified the predictions and experimental tests on suspended optical cavities \cite{zhaoPRA2008, zhaoPRA2011, KellsMod, Evans} demonstrated three-mode interactions below the instability threshold.

Three-mode interactions mimic a two-level atomic system\cite{Grudinin}, in which parametric instability is analogous to the creation of a phonon laser. Bahl et al.\cite{Bahl} emphasise that the phenomenon is a macroscopic realisation of Brillouin scattering \cite{Brillouin1965}. Because the gain depends on the product of one mechanical and two optical quality factors, the gain can be high and in principle can be tuned to cause either mode heating (leading to instability) or mode cooling \cite{zhaoPRL2009}. Both the injected mode and the scattered mode are resonant in the cavity. The circulating power of both optical modes are thus enhanced by cavity resonance, so the input power required for parametric instability is reduced. This result can be compared to two-mode parametric interactions which require detuning from their resonance and higher input power \cite{Kippenberg2010}.

The challenge to observe the phenomenon in a free space optical cavity coupled to a mechanical resonator can be met either by using very high optical power in large scale optical cavities, or at low power in tabletop cavities with suitable mode structure and a low mass high quality factor mechanical resonator. A free space cavity with an intracavity membrane creates a suitable optical mode structure which in principle can be matched to the mechanical mode structure of the membrane.

It has already been shown that silicon nitride membranes have sufficient transparency to be used in high finesse optical cavities, and have suitable high quality factor mechanical modes\cite{Zwickl2008} in the MHz range.  Well understood two-mode parametric instabilities have been observed in a suspended cavity \cite{MIT1}, and in various solid state optical microresonantors \cite{Kippenberg1, Kippenberg2, Kippenberg3, Carmon, Carmom2007, Ma}. Three-mode instability has been observed in relatively low quality factor solid state resonators\cite{Grudinin2009, Tomes2009, SAW2009, SAW2011, Kippenberg4} and in a microwave system \cite{Tobar1}. To enable three-mode parametric interactions in a free space optical cavity containing a membrane, the system must be tunable such that it sustains a pair of optical modes spaced by a frequency equal to a chosen mechanical mode frequency of the membrane \cite{Braginsky2001, Juli3mode}. The mode shapes of the mechanical modes must also be well matched in spatial structure to the  accessible cavity modes. The work present here represents the first realisation of a low loss three-mode opto-acoustic parametric amplifier.

In this paper, we first present a large amplitude model of three-mode interactions and predictions of the time evolution of a mechanical mode amplitude and an optical cavity mode amplitude. It reveals a regime in which the mechanical mode is amplified by the negative effective resistance of the cavity system, and the transition to a regime where the amplitude grows exponentially until it reaches a saturation value. The cavity design and tuning techniques are described, followed by time domain observations of the cavity high order mode amplitude. The experimental results confirm the large amplitude model, and reveal that the instability does not lead to the loss of cavity locking. The implications for advanced gravitational wave detectors are discussed.

{\it Large amplitude model of 3-mode parametric instability---} All prior analysis of three-mode interactions has assumed \cite{Braginsky2001,Kells,Braginsky2002,Juli3mode,Strigin, Evans} small amplitude. While appropriate for obtaining instability criteria this approach fails when the instability threshold is passed and the loss of fundamental mode power through scattering into high order mode becomes large.

\begin{figure}[h!]
\centering
\includegraphics[width=0.4\textwidth]{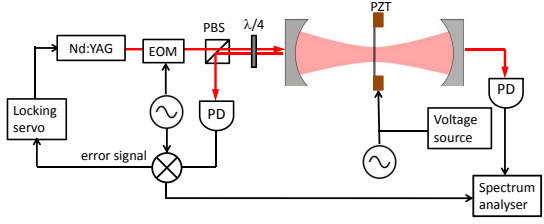}
\caption {(Color online) Three-mode interaction cavity with central silicon nitride membrane, of dimension $1mm\times1mm\times50nm$. An electro-optic phase modulator (EOM) and polarised beam splitter (PBS) lock the cavity fundamental mode frequency using Pound-Drever-Hall (PDH) locking \cite{PDH}. The membrane is mounted on the piezoactuator (PZT) to tune its horizontal position. The transmitted light from the cavity is monitored by an offset photodiode (PD) in order to detect the beat note between the fundamental mode and high order mode.}
\label{setup}
\end{figure}

We develop our theory for the experimental system shown in figure \ref{setup}, consisting of an optical cavity and a membrane in its centre, where two optical modes couple to a mechanical mode via radiation pressure. This system can be described by the interaction Hamiltonian
$
  \mathcal{H}_{\rm int} = -\hbar\lambda_{\rm OM}x_m(a_0^*+a_S^*)(a_0+a_S)\,
$
where $x_m$ stands for membrane displacement, $a_0$ and $a_S$ are TEM00 and Stokes optical modes complex amplitudes, respectively, $\lambda_{\rm OM}$ is the actual three-mode coupling strength in the experiment. Ideally, $\lambda_{\rm OM} = \Lambda\sqrt{\omega'_{0}\omega'_{S}} $ , where

\begin{equation}
\label{eq:overlap}
\Lambda = \frac{L(\int f_{0}(\vec{r}_{\bot})f_{S}(\vec{r}_{\bot})u_{z}dS}{V}
\end{equation}

is a dimensionless spatial overlapping factor \cite{Braginsky2001} of the mechanical mode and the two optical modes represented by normalised respective transverse spatial distributions $u_z(\vec{r}_{\bot})$ and $f_{0,S}(\vec{r}_{\bot})$ with $\vec{r}_{\bot}$ denoting transverse coordinates in membrane plane, and $\omega'_{0,\,S}$ stand for fundamental and Stokes cavity mode frequency shifts per unit displacement of the membrane. The latter depend on membrane reflectivity and position $x$ with respect to the cavity mirrors and equal to derivatives of modes frequencies over $x$ evaluated at the membrane rest location $x_0$:
$\omega'_{0,S}(x) = \partial_x\Bigl[(c/L)\cos^{-1}[|r_m|\cos(4\pi x/ \lambda)]\Bigr]|_{x=x_0}\,.$
Here $r_m$ is membrane amplitude reflectivity, $\lambda_0$ is laser wavelength and $L$ is cavity length.
Mode frequency dependence on membrane position is calculated using the formula for cavity resonance frequency as a function of membrane position derived in \cite{membrane2008}.

%$u_{z}$ is the vertical displacement of the membrane mechanical mode along the cavity optical axis,

The full Hamiltonian for our system reads:
\begin{equation}\label{3modeHam}
  \mathcal{H} = \hbar\omega_{0} a^*_{0}a_{0} + \hbar\omega_{S} a^*_{S}a_{S} + \hbar\omega_m b_m^*b_m + \mathcal{H}_{\rm int} + \mathcal{H}_{drive}\,.
\end{equation}
Here the first three terms describe the oscillatory dynamics of two optical and one mechanical mode, where $b_m$ stands for dimensionless complex amplitude of the membrane motion defined as $x_m = x_q(b_m+b_m^*)$ with the zero-point motion amplitude given by $x_q = (\hbar/(2m\omega_m))^{1/2}$.  Laser pump is described by the last term $\mathcal{H}_{drive} = \hbar\sqrt{2\kappa_0} A_p(a_0e^{i\omega_pt}+a^*_0e^{-i\omega_pt})$ in which $\kappa_0$ is the TEM00 mode linewidth and $A_p = \sqrt{P_{in}/(\hbar\omega_p)}$ and $\omega_p$ are the  pump laser drive amplitude and frequency, respectively, $P_{in}$ is laser power.

For slowly varying mode amplitudes, {\it i.e.} making the following transformation $b_m \to b_me^{-i\omega_m t}$, $a_0 \to a_0 e^{-i\omega_0 t}$ and $a_S \to a_S e^{-i\omega_S t}$, one can write down equations of motion as follows (equations for conjugate amplitudes can be obtained by applying complex conjugation to the equations below):
\begin{widetext}
\begin{eqnarray}\label{eq:PIfull_eqs}
  & &\dot a_0 + \left(\frac{\kappa_0}{2}-i\lambda_{\rm OM}x_q[b_me^{-i\omega_m t}+b_m^*e^{i\omega_m t}]\right)a_0 - i\lambda_{\rm OM}x_qa_S(b_m+b_m^*e^{2i\omega_m t}) = i\sqrt{\kappa_0}A_p\,,\\
  & &\dot a_S + \left(\frac{\kappa_S}{2}-i\Delta-i\lambda_{\rm OM}x_q[b_me^{-i\omega_m t}+b_m^*e^{i\omega_m t}]\right)a_S - i\lambda_{\rm OM}x_qa_0(b_me^{-2i\omega_m t}+b_m^*) = 0\,,\\
  & &\dot b_m + \frac{\gamma_m}{2}b_m - i\lambda_{\rm OM}x_q(a_0a_S^*+(|a_0|^2+|a_S|^2)e^{i\omega_m t}+a_0^*a_Se^{2i\omega_m t}) = 0\,,
  \end{eqnarray}
\end{widetext}
with detuning $\Delta = \omega_0-\omega_S-\omega_m$, $\kappa_{S}$ Stokes mode linewidth and $\gamma_m = \omega_m/Q_m$ membrane oscillations decay rate.

If we drop all the oscillating terms in the above equations and recall that pumping laser is locked to the TEM00 mode via a feedback loop with bandwidth much lower than mechanical oscillation frequency, we get the standard linearised equations of PI derived in \cite{Braginsky2001} with TEM00 amplitude equal to $A_0 = 2i\sqrt{P_{in}/(\hbar\omega_p\kappa_0)}$:
 \begin{eqnarray}
   & & \dot a_S + \frac{\kappa_S}{2}a_S - i\lambda_{\rm OM}x_qA_0b_m^* = 0\,,\\
  & &\dot b^*_m + \frac{\gamma_m}{2}b^*_m + i\lambda_{\rm OM}x_qA^*_0a_S= 0\,.
\end{eqnarray}
Thereof one can immediately derive a parametric instability condition, assuming solution in the form $\{a_s,b_m^*\}\propto e^{\Gamma t}$ and requiring $\Gamma\geqslant0$:

\begin{equation}\label{eq:PIcond}
\mathcal{R}\equiv\frac{8\lambda^2_{\rm OM} P_{in}}{m\omega_m\omega_p\gamma_m\kappa_0\kappa_S}\geqslant 1\,,
\end{equation}

The threshold value for input laser power at which PI arises can be calculated by rearranging this equation,
\begin{equation}
\label{eq:Pin_thres}
P_{in}^{thres} = \frac{m\omega_m^2\omega_p^3}{8\lambda^2_{\rm OM}Q_{m}Q_{0}Q_{s}}
\end{equation}

The ringup time, with the assumption $\gamma_m\ll\kappa_S$, reads:
\begin{equation}\label{eq:ringup_theory}
  \tau_{r} = \Gamma^{-1} \simeq \frac{m\omega_m\omega_p\kappa_0\kappa_S}{4 \lambda^2_{\rm OM}P_{in}}\,.
\end{equation}

To explain the saturation phenomenon we need to account for the omitted oscillating terms in \eqref{eq:PIfull_eqs}. It can be done by using the combination of the method of harmonic balance and the method of slowly varying amplitudes. We look for a solution in the form $A(t) = \sum_{n=-\infty}^\infty\alpha_n(t)e^{-in\omega_m t}$ where $A(t)$ refers to the optical modes, $a_0$, $a_S$ and $\alpha_n(t)$ denoting the slowly varying amplitudes of the corresponding mode harmonics (slow compared with $\omega_m^{-1}$). Substituting the above solutions into the initial equations \eqref{eq:PIcond} and doing lengthy but straightforward calculations, we obtain the following equation for the mechanical amplitude:
\begin{equation}\label{eq:PI_slow_b}
  \dot b_m + \frac{\gamma_m}{2}\left(1-\mathcal{R}\frac{J^2_0\left(k_1b_m\right)G^2_0G^*_S}{|1+k_2 G_0 G_S|b_m|^2|^2}\right)b=0
\end{equation}
 where $G_0 = 1-e^{-\kappa_0 t/2}$, $G_S = 1-e^{-(\kappa_S/2-i\Delta)t}$, $J_0(x)$ is the Bessel function of the order 0, $
k_1 = 2\lambda_{\rm OM} x_q/\omega_m
$,
$
k_2 = (2\lambda_{\rm OM}x_q/\sqrt{\kappa_0\kappa_S})^2(1-2i\Delta/\kappa_S)
$. One can also derive expressions for the amplitudes of harmonics of both optical modes:
 \begin{eqnarray}
   & & \alpha_{0,n} = J_n(k_1 b_m)\frac{A_0G_0J_0(k_1b_m)}{1+k_2 G_0 G_S|b_m|^2}\,,\label{eq:PI_slow_a0}\\
   & & \alpha_{S,n} = -k_3J_n(k_1 b_m)\frac{A_0G_0G_SJ_0(k_1b_m)}{1+k_2 G_0 G_S|b_m|^2}\label{eq:PI_slow_aS}\,,
 \end{eqnarray}
 with $k_3 = 2\lambda_{\rm OM}x_q/\kappa_S$. We see that the optical mode dynamics is fully governed by that of the mechanical mode. We can also derive a steady-state amplitude of the mechanical mode from Eq.~\eqref{eq:PI_slow_b}, setting $\dot b_m = 0$ and solving the resulting non-linear algebraic equation with respect to $b_m$ taking into account that $G_{0,S}(t\to\infty) = 1$.

\begin{figure}[h!]
\centering
\includegraphics[width=0.5\textwidth]{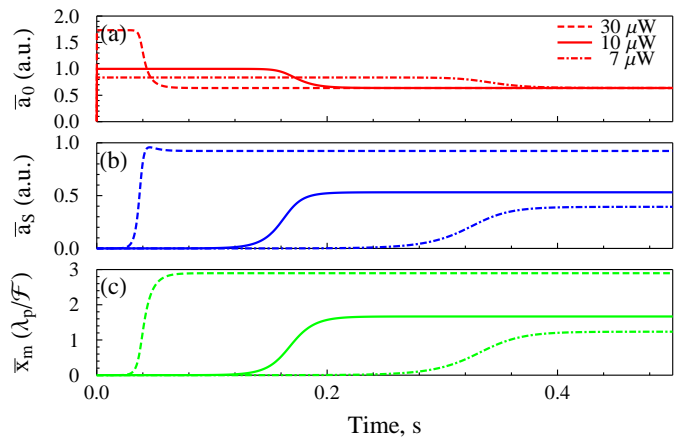}
\caption {(Color online) Theoretical model for the time evolution of parametric instability. (a) the cavity fundamental mode; (b) the cavity high order mode; and (c) mechanical mode. Optical modes are normalised by the value of $|A_0|$ at $P_{in}=10\ \mu\mathrm{W}$. The mechanical mode amplitude is given in units of Fabry-P\'erot cavity linear dynamic range. (Here $\lambda_p$ is the pump laser wavelength and $\mathcal{F}$ is the cavity finesse.)}
\label{fig:theory}
\end{figure}

The resulting dynamical evolution is given in figure~\ref{fig:theory}. We see that as the Stokes mode and the mechanical mode ring up, the circulating power in the TEM00 mode goes down. When the mechanical oscillation amplitude approaches the Fabry-P\'erot cavity linear dynamic range, $\lambda_p/\mathcal{F}$, the system reaches saturation and all modes acquire steady amplitudes that depend on input laser power \cite{Polyakov}. Further detailed analysis of  parametric instability in the highly non-linear regime will be published elsewhere \cite{Future}.

% \section{Optical cavity design}
{\it Experimental observation of three-mode parametric instability ---}To observe the three-mode interaction, two relevant optical modes (TEM00 and TEM02 in this experiment) must be simultaneously resonant inside the cavity. To obtain maximum gain \cite{Braginsky2001}, the frequency difference between the two optical modes must be tuned to the mechanical mode frequency.  The mechanical modes are modes of a silicon nitride membrane with frequencies spanning from 0.1MHz to a few MHz. However the free spectral range of a $~10cm$-long tabletop optical cavity is in the range of GHz.  The mode spacing needs to be tuned with a precision of $\sim 10^{-6}$ of the free spectral range. This can be achieved  with careful choice of  mirror radii of curvature and cavity length. For our case, a near-confocal cavity (mirror ROC close to the cavity length) is tuned so that the frequency of the higher order TEM02,p mode close to the frequency of the next longitudinal mode TEM00,p+1. Small length adjustments enable the resonant condition to be met. In the cavity with a central silicon nitride membrane shown in figure \ref{setup}, the membrane position provides a second way of tuning the cavity mode frequencies. The cavity resonance frequencies with the membrane in the middle, compared to the empty cavity, are shifted by an amount determined by the reflectivity and position of the membrane \cite{membrane2008}, given by
$
 \omega_{cav} = (c/L)\cos^{-1}[|r_m|\cos(4\pi x_m/ \lambda)]\,,
$
where $x_m$ is the membrane position relative to the centre of the cavity.
\begin{figure}[!b]
\centering
\includegraphics[width=0.4\textwidth]{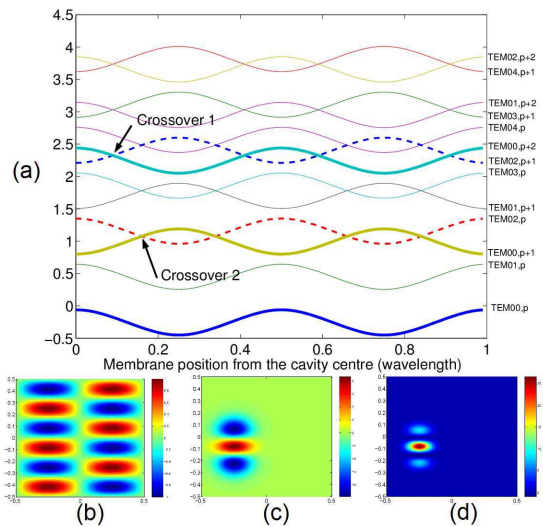}
\caption {(Color online) Cavity mode frequency tuning, mode shapes and overlap. (a) mode frequencies as a function of membrane position. (b) The mode shape of the membrane(2,6) mode; (c) mode shape of the TEM02 cavity mode. The product of the three mode shapes (d) confirms the strong mode shape overlap requested to obtain the three-mode interaction.}
\label{freq}
\end{figure}
Figure \ref{freq} shows $\omega_{cav}$ for different cavity modes as a function of membrane position. When the membrane is within a quarter of the wavelength from the centre of the cavity, there are two frequencies crossovers between the TEM00 mode and the TEM02 mode, labeled as "crossover 1" and "crossover 2". Tuning the membrane position in principle allows the mode spacing between the TEM00 and TEM02 modes to be matched to the membrane mechanical mode frequency. In practice, the crossovers  are avoided  due to coupling between the modes \cite{OptLett, HarrisTuning}. Normally there exists a minimum frequency spacing at the nominal crossing point. The greatest challenge in this experiment was the adjustment of the membrane position and alignment to allow the minimum frequency spacing to be close to the chosen mechanical mode frequency.

% \begin{figure}[!b]
% \centering
% \includegraphics[width=0.4\textwidth]{modes.eps}
% \caption {(Color online) A set of modes: (a) Membrane mode (2,6); (b) Cavity second order mode TEM02; and,(c) Product of three modes, TEM00, TEM02 and membrane mode (2, 6).}
% \label{fig:ex3}
% \end{figure}

The second challenge for a three-mode interaction experiment is to design a significant spatial overlap between the mechanical mode and the optical modes. Figure \ref{freq} shows the amplitude distribution of one membrane mechanical mode denoted (2,6), the optical cavity TEM02 mode, and the product of all three modes ((2,6), TEM02, TEM00). It is clear that there is a good overlap between these modes if the cavity modes are correctly positioned on the membrane at a specific location which is not at the centre of the membrane. The calculated maximum overlap factor is $\sim0.8$.

% \section{Cavity setup and Tuning}
% The cavity mirrors are mounted on motorized optical mounts which have angular range of $ \pm4 \,^{\circ}$ about $x$ axis (pitch) and $y$ axis (yaw) respectively. The membrane is made of silicon nitride of 50nm thick, and 3mm by 3mm square. It is attached to a piezoactuator providing both high frequency excitation and $500nm$ tuning range along $z$ axis (along the cavity optical axis). The piezoactuator is mounted on a motorized multi-stage optical mount to provide angular adjustment as well as $10mm$ movable range along x, y and z axis. All the optical mounts are fixed on a bar made of Invar, placed in a vacuum tank. The tank and all other optical components are mounted on an air-floated table to isolate seismic noise. The tank is pump down to $10^{-4}mbar$ by a turbo molecular pump. Special care is taken to isolate vibration transfer from pump to the tank. However there is still some residual low frequency noise coupled into our system.

Our optical cavity was mounted on an invar bar in a vibration isolated vacuum tank. Motorised optical mounts and piezoactuators were used for cavity alignment. (Details available from the authors). The experimental setup is as shown in figure \ref{setup}. We developed careful alignment and tuning procedures to tune the cavity. First the cavity finesse $\sim15000$ was measured without a membrane present. Then the membrane, which had been previously aligned using a He-Ne laser, was inserted into the cavity. The maximum finesse observed with the membrane inserted was $\sim13000$ (photon lifetime is $1.38 \mu s $).

% After rough alignment of the mirrors and membrane using a He-Ne laser, the membrane is translated out of the cavity to allow initial tuning of the cavity TEM00 mode using photodetectors to monitor both transmission and reflection and a CCD camera to identify  mode shapes.  The cavity finesse of $\sim15000$ was measured by monitoring the ring down time of the transmission beam. After the initial alignment the membrane is re-inserted  at  the middle of the cavity. The maximum observed finesse in this condition was $\sim13000$.

% \begin{figure}[b!]
% \centering
% \includegraphics[width=0.48\textwidth]{membraneyawtuning.eps}
% \caption{The cavity mode spacing between TEM00 mode and TEM02 changes as the membrane misalignment.}
% \label{fig:membranetuning}
% \end{figure}

Tuning the optical mode frequency spacing was achieved by tuning the membrane position and orientation. We chose "crossover 1" labeled in figure \ref{freq} as the target because the membrane position happens to be close to a node of the TEM02 mode. In this location the membrane absorption is minimum, allowing the highest possible cavity finesse. We measured the dependence of TEM00 and TEM02 mode frequency spacing on membrane angle. The frequency spacing should be minimised when the membrane is normal to the incident light and was tuned at a rate $\sim4MHz/ mrad$. We tuned the membrane position along the optical axis to the desired "crossover 1" position, while the gap spacing at the avoided crossing was tuned by membrane angular adjustment.

The measured membrane mode resonant frequency agrees with the calculated value of $f_{i,j} = \sqrt{T/4 \rho d^2}\sqrt{i^2+j^2}$ with membrane tension $T = 800MPa$ and density $\rho = 2.7 g/cm^3$. The indices $i$ and $j$ are the mode numbers. However the membrane frequency drifted over time. For example the (1,1) mode dropped from $\sim402kHz$ to $\sim384kHz$ over 1 year. At the time of the measurements reported here the (2,6) mode frequency was $\sim1718kHz$. At pressure $10^{-3}mbar$, the membrane has a Q-factor of $10^5$, corresponding to a phonon lifetime of $ 10 ms $. % The maximum Q factor measured with out optical pump for membrane 26 mode ($1718kHz$) is $ $. }

% The pre-amplified signals were fed to a spectrum analyser to observe membrane mechanical mode amplitude. The spot size on the membrane is $\sim0.3mm$, but it is difficult to determine the exact spot size on the membrane as very little light is scattered from the membrane. The incident optical power must be kept very low since parametric instability occurs for only a few $\mu W$ power injected into the cavity. The TEM02 signal is below electronic noise when the injected power is lower than $1 \mu W$, so that we can only detect the mechanical mode signal when the instability occurs and the mechanical mode amplitude builds up.

For the very low mass membrane and the high finesse cavity used in this experiment, the threshold for the parametric instability is $\sim 5 \mu W$. Hence the experiments had to be conducted at very low optical power. This means that the TEM02 mode is below the photo detector noise floor except when the mechanical amplitude has built up. The TEM02 mode was detected with a simple, partially shadowed photodetector since a quadrant photodetector was not available.

\begin{figure}[h]
\centering
\includegraphics[width=.48\textwidth]{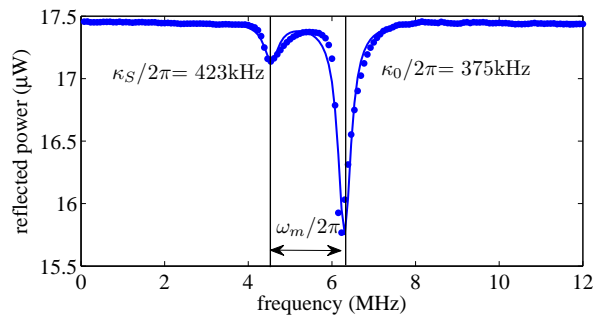}
\caption {(Color online) The cavity mode frequency spectrum when the mode spacing is tuned to the mechanical mode frequency.}
\label{gap}
\end{figure}

\begin{figure}[h]
\centering
\includegraphics[width=0.48\textwidth]{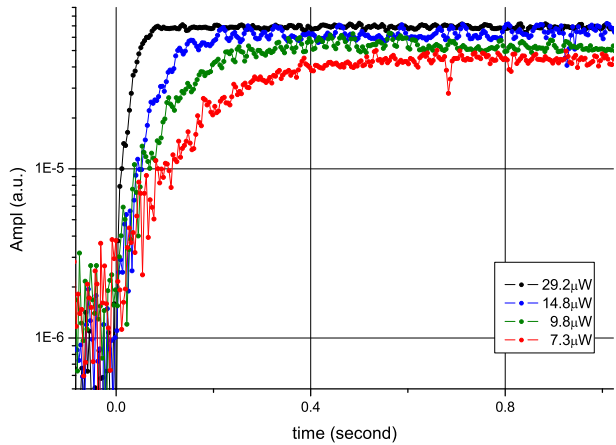}
\caption {(Color online)The photodetector signal as a function of time at different input power level, which shows exponential ringup. This signal measures the beat note between the cavity TEM00 mode and TEM02 mode, and is an indirect readout of the mechanical mode amplitude.}
\label{fig:ex4-a}
\end{figure}

\begin{figure}[h]
\centering
\includegraphics[width=0.48\textwidth]{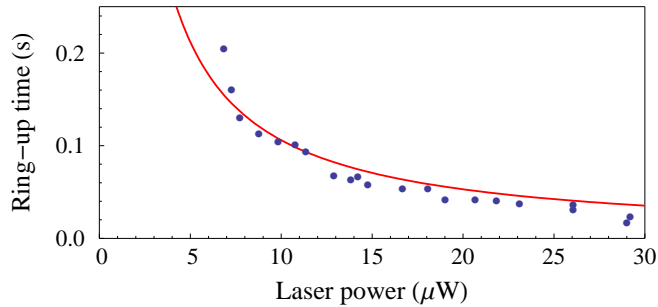}
\caption {(Color online)The photodetector signal ring-up time as a function of the input power. The solid curve is the curve fitting to the theoretical model.}
\label{fig:ex4-b}
\end{figure}

% \section{Experimental Results}
Using the tuning method described above we tuned the optical mode spacing to the (2, 6), 1718kHz membrane mode as shown in figure \ref{gap}. As the optical mode spacing decreases, the cavity mode linewidth increases due to the coupling between the two modes \cite {OptLett}. We fitted the tuning curve in the figure \ref{gap} to two Lorentzians to determine the corresponding mode linewidth. These cavity mode linewidths were used in the modeling results given in figure \ref{fig:theory}.  The cavity stayed tuned  long enough to make repeatable measurements, but frequency drift meant that retuning was required everyday.

Once correctly tuned, exponential ring up of the (2,6) mechanical mode occurs whenever the input power exceeds the threshold. Figure \ref{fig:ex4-a} shows the Stokes mode amplitude as a function of time at various input powers. Clearly higher input power corresponds to faster ring-up as expected. Figure \ref{fig:ex4-b} shows the ring-up time as a function of the input power. The ring-up time decreases with input power in close agreement with the theoretical model. The saturation is clearly visible in figure~\ref{fig:ex4-a}. The mechanical amplitude reaches the cavity linewidth $\sim10^{-10} m$ in a time between 0.1 and 0.5 seconds, in agreement with theory. The dependence of saturation amplitude on input power is also visible in the figure~\ref{fig:ex4-b}. Using \eqref{eq:PIcond} and \eqref{eq:Pin_thres}, we can estimate an optomechanical coupling strength for our system as $\lambda_{\rm OM}/2\pi= 0.84\times10^{14} \mathrm{Hz/m}$ and a threshold laser power for parametric instability of $P_{in}^{thres} = 5.7\ \mu\mathrm{W}$.

% As one can see from figure~\ref{fig:ex4-a} the amplitudes of Stokes mode (mechanical mode) oscillations reach a saturation threshold. This phenomenon is due to the limitation on the mechanical oscillation amplitude set by the high-finesse ($\mathcal{F}\simeq 10^4$) cavity linewidth. Once the mechanical oscillation amplitude reaches the value of the cavity linewidth in displacement, it introduces time-dependent linear detuning $\propto \lambda_{\rm OM}x_m(t)$ of the cavity from the pump laser frequency. This results in a decrease of the optical power stored in the TEM00 mode and through this sets the limitation on the power available for mechanical and Stokes mode oscillations ring-up.

% \section{Conclusion}
{\it Conclusion ---}We have shown that the theory of  three mode opto-mechanical interactions correctly predicts the onset of parametric instability in a three-mode opto-mechanical system. The results are in excellent quantitative agreement with the original theory of Braginsky. We find that mechanical mode amplitudes saturate in accordance with the large signal model for parametric instability presented here. In the system studied here, the onset of three mode instability does not lead to loss of locking of the main optical cavity. The loss of power from the main cavity mode is sufficient to stabilise instability. If the same behavior is observed in high power laser interferometers for gravitational wave detection, it should be much easier to implement instability control techniques based on feedback or slow thermal tuning.

{\it Acknowledgment} This research was supported by the Australian Research Council. We would like to thank the referees for helping us improve this manuscript. We wish to thank the Gingin Advisory Committee of the LIGO Scientific Collaboration, the LIGO Scientific Collaboration Optics Working Group and our collaborators Pierre-Francois Cohadon, Antoine Heidmann, Stefan Gossler, Gregg Harry and Stan Whitcomb for encouragement and useful advice.

%\bibliographystyle{unsrt}
%
%\bibliography{F:/myreference/myreference}

\end{document}